\documentclass[a4paper,margin=1in]{article}
\usepackage{geometry}
\usepackage{multicol}
\usepackage[utf8]{inputenc}
\usepackage{mathtools}
\usepackage{amsthm}
\usepackage{amssymb}
\usepackage{hyperref}
\usepackage[usenames,dvipsnames]{xcolor}
\usepackage{todonotes}
\usepackage{algorithm}
\usepackage{algpseudocode}
\usepackage[toc,page]{appendix}
\usepackage{tikz}
\usepackage{esvect}
\usepackage{xcolor}
\usepackage[T1]{fontenc}
\usepackage{lmodern} 
\usepackage[numbers]{natbib}
\usepackage{booktabs}
\usetikzlibrary{patterns}
\usetikzlibrary{decorations.pathreplacing}

\usepackage{graphicx}
\graphicspath{ {img/} }

\DeclareUnicodeCharacter{00A0}{ }
\usepackage{hyperref}
\hypersetup{
    colorlinks = true,
    linkbordercolor = green,
    pdfborder = true,
    citecolor= blue,
    pdfauthor={Andrzej Pacuk, Piotr Sankowski, Karol Wegrzycki, Piotr Wygocki},
    pdfsubject={There is something beyond the Twitter network},
    allbordercolors=green,
    linkcolor = Red,
}

\newcommand{\const}{\ensuremath{\mathrm{const}}} 

\newcommand{\footremember}[2]{%
    \footnote{#2}
    \newcounter{#1}
    \setcounter{#1}{\value{footnote}}%
}
\newcommand{\footrecall}[1]{%
    \footnotemark[\value{#1}]%
} 
\title{There is something beyond the Twitter network}
\author{Andrzej Pacuk\footremember{MIMUW}{Institute of Informatics, University
    of Warsaw, Poland} \and Piotr Sankowski\footrecall{MIMUW} \and Karol W\k{e}grzycki\footrecall{MIMUW} \and Piotr
Wygocki\footrecall{MIMUW}}
\date{\texttt{[apacuk,sank,k.wegrzycki,wygos]@mimuw.edu.pl}}
\begin{document}

\maketitle

\begin{abstract}

How information spreads through a social network? Can we assume, that the
information is spread only through a given social network graph?  What is the
correct way to compare the models of information flow? These are the basic
questions we address in this work.

We focus on meticulous comparison of various, well-known models of
rumor propagation in the social network. We introduce the model incorporating mass
media and effects of absent nodes. In this model the information appears
spontaneously in the graph. Using the most conservative metric, we showed that
the distribution of cascades sizes generated by this model fits the real data much
better than the previously considered models.

\end{abstract}

\section{Introduction}

Sociology, empirical investigation, critical analysis, social policy, political
science and market analysis---it is just the beginning of the long list
of research areas focused on the information diffusion. Today, when
more and more communication can be tracked and logged, those fields of 
research can benefit from the analysis of the social networks.
The research on the information diffusion can be
applied to maximize the influence~\cite{max-influence} and virality of the rumor,
enhancing recommender systems or improving routing
algorithms~\cite{iwanicki}.

Current, state-of-the-art models of the information diffusion do not correctly
predict the distribution of cascades sizes~\cite{cikm2014} (the cascade size is
the number of nodes sharing a given information). Application of these
models in the aforementioned areas can lead to a hardware overload,
non-optimal recommendations erroneous predictions about influence.  In this
work we claim that, the main problem is the fact that the
information does not diffuse only in the known social network. Namely, there is
something beyond the easily observable relations.  Our analysis on the Twitter network
has showed that the information spreads not only by the known connections of
acquaintance. To model the information diffusion one needs to
incorporate effects of inoculating a rumor by absent nodes, different social
networks, mass media or a word of mouth. All effects might be
responsible for spawning the rumor by nodes that are not connected in social network.
We introduce the \emph{multi-source $\alpha^k$} model, that can be expressed as an
extension to the standard \emph{compound Poisson process}. This model
incorporates the spontaneous rumor inoculation in the simple information diffusion model.
The application of such well based method allows us to tune the cascade size
distribution to a real data with almost no cost in the time complexity.

The contribution of this work is to propose a simple, replicable metric
and an anonymised dataset to provide the measurable comparison
between information diffusion models.
Moreover, we provide the efficient source code to evaluate basic
models~\cite{approx2016}. 
Finally, we claim that the 
Kolmogorov-Smirnov test should be used to collate distributions of information
cascades sizes. Based on that test, we showed that the effects of mass media
are substantial to model the information diffusion.

\subsection{Related Work}

The dynamics of an information flow in social networks has
been studied by numerous researchers~\cite{cikm2014,greg,lerman}.

Originally, the epidemiology and the solid-state physic areas suggested different
models such as SIR (susceptible infectious recovered) or SIS
(susceptible infectious susceptible). These models had been employed to predict
dynamics of information spread. However, all of these models assume that
everyone in the population is in the contact with everyone else~\cite{bayley}, which
is unrealistic in the large social network.

The classical example of a modified spreading process has been considered in~\cite{wlosi} 
by adding the effect of \emph{stifler}. \emph{Stiflers} never spread
the information, even if they were exposed to it multiple times. Nevertheless,
\emph{stiflers} can actively convert spreaders or susceptible nodes into
\emph{stiflers}. Nevertheless, it is unclear who would act as a \emph{stifler} in
the fast news propagating network like Twitter or Digg.

\citet{leskovec-blog} proposed the cascade generation model
and simulated it on the dataset of blog links. Their model assumed that 
every connection in the information propagation graph is equally important
and used a single parameter that measures infectiousness of an average rumor. 
Then, they acknowledge
a simple improvements in their model of the cascade generation. Most notably, they noted that exponential 
decrease of the infectiousness can also be considered to model the information
spread.

Collaterally, researchers have analyzed properties of the social network
graph~\cite{Ghosh,wlosi} and the distribution of the cascades sizes (or rumor
popularities)~\cite{moro}. Today, it is well known, that cascades sizes follow a heavy
tail distribution~\cite{lerman} but due to the lack of appropriate data it is still
unclear, whether it follows the power-law (Pareto) or the lognormal distribution.

Since then, computer scientists noticed that the state-of-the-art rumor
propagation models do not
predict such distribution. \citet{greg} observed that information propagates
onto an entire graph too often and \citet{cikm2014} showed an evident phase
transition for middle sized rumors. However, those observations lack the systematized
metric to compare predictions with the real cascade size distribution. In this
work we claim that K-S test is the most conservative benchmark for testing these
features. Moreover, we propose that every experiment ought to be meticulously 
tested to avoid misinterpretations. 

\section{Modeling Information Cascades}\label{model}

\subsection{Comparison Metric and Evaluation}\label{eval}

Since heavy tailed distribution of cascade sizes had been observed, there has been
many attempts to readjust the diffusion models to the real
data~\cite{cikm2014,greg}. Unfortunately, none of them has proposed an adequate metric to
measure this distribution.

It is tempting to propose a metric that would somehow
punish errors on a tail of distribution (in contrary to exponential
distribution, the large events do happen more often in the power-law
distribution).
Another naive idea is to assume that the power-law distribution 
is linear on a log-log plot and use the linear regression to fit it.
Unfortunately, \citet{why-kstest-for-powerlaw} claim that both of those methods have 
serious problems with variation and many distributions might be misclasified
by these metrics.

The method of analyzing a power-law distributed data should~\cite{why-kstest-for-powerlaw} 
involve goodness-of-fit test. The most commonly used is Kolmogorov-Smirnov
test (K-S test):
\begin{displaymath}
    \sup_{x} | X(x) - Y(x)|
    ,
\end{displaymath}

which computes the maximal difference between cumulative distribution functions (CDF) of a real and predicted 
distributions ($X(x)$ is the CDF of a predicted data and $Y(x)$ the CDF of a real data).
Recently, \citet{ks-twitter} showed that using
the aforementioned methods, the lifetime of a tweet does not follow the Pareto distribution but in fact
it is the type-II discrete Weibull distribution.

Introducing models with multiple parameters may lead to the serious overfitting.
Because, when comparing to the Pareto distribution of cascade sizes, returnining only cascades with
size smaller than $10$ can result K-S test smaller than
$0.01$. When one introduces a flexible model like ICM~\cite{KDD03}, one needs to
carefully analyze the model to avoid overfitting. It might be tempting to use
machine learning approach based on the large number of features. This might
result in a better fit to the real distribution of cascades sizes. On the other
hand, the reasoning based on this approach can be hard.
In this paper, we focus on the models that
describe the fundamental mechanisms of the information diffusion. 

The analysed models are trained using grid search. We chose the parameters,
so the empirical error of K-S test is smaller than $0.001$.

\subsection{Dataset Description}\label{dataset}

We analyzed the set of over 500 million tweets, extracted from
10\% sample of Twitter tweets collected from May 19 to May 30 2013. In our dataset, each tweet beside 
the content of the tweet contained hashtags list, user id and if existed
ids of the mentioned users, retweeted user id and id of the user this tweet was
a reply to.

Based on that, we generated the graph of retweets: the vertices in graph are
user ids that occurred in our dataset. If there exists a tweet with user $A$
retweeting, replying or mentioning user $B$, then we add an edge from vertex $A$ to
$B$. The graph of retweets contained 71 million vertices and 230 million edges. 

The popularity of the hashtag is the number of users that had used it. Recall,
that our goal is to analyze the information diffusion on social network. Hence,
we have focused merely on fresh hashtags: we keep hashtags that did not occur
during the first day of our sample. 
The specific day used as the first day for new hashtags should not influence 
the results of the tests, because the majority of the cascades last less than 
few days~\cite{Kwak}.  
Finally, we obtained 7.7 million of distinct,
fresh hashtags. Based on these hashtags, we generated popularity distribution,
that will be used to compare models of information diffusion.

The graph of retweets strongly depends on the number of gathered events.
Be aware that the parameters of the presented algorithms depend on the specific 
retweet graph. The parameters for graph produced from all events from a given month will be probably different.
For example, in the introduced models, the probability that an individual will retweet decreases with the density of the graph.
 On the other hand, we need general models
independent from the instance of the network.
A standard approach to avoid overfitting is to divide the set of tweets into
two independent subsets: the \underline{training set} used for tuning parameters
and the \underline{test set} used for the validation. We divided the set of the retweets into two sets
each containing half of the available, consecutive, full days. We evaluated 
model $\alpha$, model $\alpha^k$ and multi-source $\alpha^k$
on both training and test sets. The grid search shows that the optimal
parameters for training and test set are identical. The K-S test values for both
sets are identical up to $0.001$ error. Overall ranking of the models is not
changed. Hence, our analysis is not prone to overfitting.
To achieve better precision, through the rest of the work, we use the
graph based on all events.

To promote the study on information diffusion we share 
our dataset with the other researchers~\cite{approx-data-2016}.
We have removed unnecessary parameters and anonymized our dataset according
to the Twitter rules regarding public sharing.

\subsection{The Graph of Retweets versus the Follower-Followee Graph}

There is another choice of information dissemination graph, namely, the
follower-followee graph. \citet{ks-twitter} claim, that people are more selective in what
they say, rather than whom they listen. Moreover, \citet{ks-twitter} assert that the graph of
retweets may encode the true interest among the users better than
follower-followee graph. Since the
cascades consist of interactions, the graph of retweets seems to be a better
choice than the follower-followee graph. Note that follower-followee
relationship is perturbed by a Twitter recommendation system.  Altogether, we
believe that the graph of retweets describes the relationships
between the users much better than the follower-followee graph. Moreover, the current
retweet graph can be obtained by using the Twitter API~\cite{api}. Unfortunately
the up-to-date, follower-followee graph is currently unavailable.

\subsection{Cascade Generation Model}\label{model_cgm}

The \emph{cascade generation model} (CGM) introduced by~\cite{leskovec-blog}
uses constant $\alpha$, which is the probability that the
information is passed from a user to its follower.

According to~\cite{leskovec-blog} the cascade is generated in the following steps:
\begin{enumerate}
    \item Uniformly at random pick a starting point of the cascade
        and add it to the set of \emph{newly informed} nodes.
    \item Every \emph{newly informed} node, for each of his direct neighbors,
        makes a separate decision to inform the neighbor with
        the probability $\alpha$.
    \item Let \emph{newly informed} be the set of nodes that have
        been informed for the first time in step $2$ and add them
        to the generated cascade.
    \item Add all \emph{newly informed} nodes to the generated cascade.
    \item Repeat steps $2$ to $4$ until \emph{newly informed} set
        is empty.
\end{enumerate}
In CGM regime all nodes have an identical impact ($\alpha = \const$).
The final graph of the information spread is called a cascade.

\subsection{Model Alpha}\label{model_alpha}

CGM is modeling communication with all connected nodes independently. That is, in the step $2$, newly informed
node might potentially pass the information only to some subset of its acquaintances.
However, Twitter is a microblogging platform where messages posted by a user are instantly
received by all of its followers. Then, each follower may share these messages with his followers
by replying, retweeting or mentioning.

We propose the \emph{model $\alpha$} which resembles the CGM, but is better
suited for the schemes of communication in the Twitter network.
The single cascade of \emph{model $\alpha$} is generated as follows: 
\begin{enumerate}
    \item Uniformly at random pick a starting point of the cascade
        and add it to the set of the \emph{newly informed} nodes.
    \item Every \emph{newly informed} node independently with
        the probability $\alpha$ becomes a \emph{spreader} and then informs all their
        direct neighbors.
    \item Let \emph{newly informed} be the new set of nodes that have
        been informed for the first time in step $2$.
    \item Add all new \emph{spreading} nodes to the generated cascade.
    \item Repeat the steps $2$ to $4$ until the \emph{newly informed} set
        is empty.
\end{enumerate}

Indeed, the \emph{model $\alpha$} differs from CGM:
\begin{itemize}
    \item The main difference is in the point $2.$, where in the \emph{model
        $\alpha$} the newly informed node makes a single decision: either
        to inform all of his followers or to inform none of them.
    \item In the CGM nodes might had multiple chances to become a spreader (after
        receiving an information from each of the followed nodes).  However, in
        the model $\alpha$, each of the informed nodes has just one chance to become
        a spreader: only after being informed for the first time.
    \item The final cascade size is the total number of spreaders, whereas in CGM
        it is the number of informed nodes.
\end{itemize}

\subsubsection{Experimental Results}

We simulated model $\alpha$ on the graph of retweets.
We used \emph{grid search} algorithm with step $0.0001$ to tune parameter
$\alpha$. Subsequently, for every $\alpha$
we ran 10 million simulations to generate the cascades size distribution.
Then we computed K-S test for each distribution.
The best K-S test $0.0447$ was achieved for $\alpha=0.0884$.
Our experiments show that roughly $4\%$ of cascades in this model are larger
than $10\,000$. On the other hand, in the real data, the large cascades
constitute $0.01\%$ of all cascades.
This amount of the large cascades is the main reason for such low K-S test
result of the model $\alpha$. Hence, we need a model in which the number of
extremely large cascades is heavily reduced.



\subsection{Exponential Model}\label{model_alpha_k}

Twitter rumors have a limited lifetime.
The information obeys the effect:
the further from source, the lower the virality of the information.
\emph{Model $\alpha^k$} will simulate that process by decreasing
the infectiousness of the information after each
round. The only difference from \emph{model $\alpha$} is decreasing
probability of becoming a spreader:

\begin{itemize}
    \item In the first round each neighbor of a initial vertex is informed and then with probability $\alpha$ becomes the spreader.
    \item During the round no. $k$ each previously, not informed neighbor of the new
          spreaders from the round $k-1$ is informed and subsequently, with probability $\alpha^k$ becomes a spreader.
\end{itemize}

\subsubsection{Experimental Simulations of Exponential Model}

Once again, we conducted simulations on Twitter retweets graph.
We used \emph{grid search} algorithm with step $0.0001$ to tune parameter $\alpha$.
For each $\alpha$ we performed $10$ million simulations and
then we computed K-S test on each distribution versus real distribution.
For $\alpha=0.1357$ we obtained the best K-S test value $0.0207$ which is
roughly the half of the value of model $\alpha$.
The K-S test value is mostly affected by the probabilities of a few smallest
cascades sizes.
Namely, in \emph{model $\alpha^k$} $74.2\%$ of cascades have size $1$, versus $76.2\%$ in the real data. 



\subsection{Multi-source Exponential Model}\label{model_multisource_alpha_k}

The exact structure of the connections between all people in
the world is unknown. It can be modeled by the graph of followers~\cite{brach2}
or the graph of retweets~\cite{cikm2014}, which can successfully approximate
the real connections. Because of existence of mass media, absent nodes or
communications through channels unavailable for researchers such as telephones,
private conversations or emails, sometimes the information emerges
somehow randomly in the new source nodes.

We have a graph with extremely many nodes $n$. Moreover, probability $p$ that a randomly
selected node will spread the information it was not informed of, through our network is
ridiculously low. Hence, the number of spreaders that get to known the information
from a different source can be modeled by the Binomial distribution:

\begin{displaymath}
    X \sim \operatorname{B}(n,p)
    .
\end{displaymath}

However, by the law of rare events, this can be approximated by Poisson
distribution:

\begin{displaymath}
    X \sim \operatorname{Pois}(np)
    .
\end{displaymath}

Of course here, we assume that we do not consider globally known rumors, where probability $p$
is not that low (e.g., the information concerning a soccer match TV transmission,
where Twitter users share the exceptional achievements through a social
network).

We can model the information diffusion as follows:
\begin{itemize}
    \item Randomly choose the first node that will be informed.
    \item Propagate the information using the \emph{model $\alpha^k$} from the previous section.
    \item Until there are new, informed nodes, in each round randomly choose
        $X \sim \operatorname{Pois}(\lambda)$ new source nodes and
        propagate information from those nodes by \emph{model $\alpha^k$}.
\end{itemize}

This model has one additional parameter $\lambda$ which is interpreted as
the expected number of nodes randomly informed in each round.
\citet{leskovec-blog} show that, the single-sourced cascades very rarely
collide with each other. Hence, if one assumes that all cascades are
disjoint, then the final rumor size can be approximated by the sum of the
cascades sizes. Procedure
\textsc{MultiSourceAlphaExp($\alpha, \lambda$)} (see Algorithm
\ref{multisource_alpha_k_algorithm}) computes the size of random rumor, where:
\begin{itemize}
    \item \textsc{RandomAlphaExpCascade($\alpha$)} returns properties (the total size and the number of rounds)
          of a random cascade generated by \emph{model $\alpha^k$}.
    \item \textsc{RandomPoisson($\lambda$)} gives a random integer from $\operatorname{Pois}(\lambda)$.
\end{itemize}

The distribution of cascade properties (pairs of the total
size and the number of rounds) generated by
the \emph{model $\alpha^k$} can be precomputed. So choosing the random cascades
will take $O(1)$ time. Hence, the time complexity of Algorithm~\ref{multisource_alpha_k_algorithm}
is $O(t(\lambda+1))$, where $t$ is the expected number of rounds.
However, for sufficiently large values of $\lambda$ the algorithm may not stop.

\begin{algorithm}
\caption{Model multi-source $\alpha^k$}
\label{multisource_alpha_k_algorithm}
\begin{algorithmic}
\Procedure{MultiSourceAlphaExp}{$\alpha, \lambda$}
\State $(\mathrm{size}, \mathrm{rounds}) \gets  \Call{RandomAlphaExpCascade}{\alpha}$
\For{$\mathrm{curr\_round} \gets 1 \text{ to } \mathrm{rounds}$}
    \For{$i \gets 1 \text{ to } \Call{RandomPoisson}{\lambda}$}
        \State $(s, r) \gets  \Call{RandomAlphaExpCascade}{\alpha}$
        \State size $\gets \mathrm{size} + s$
        \State rounds $\gets \max(\mathrm{rounds}, \mathrm{curr\_round} + r)$
    \EndFor
\EndFor
\State \textbf{return} size
\EndProcedure
\end{algorithmic}
\end{algorithm}

Now, the results of multi-source $\alpha^k$ model can be reformulated as:
\begin{displaymath}
    X_0 + Y(t) = X_0 + \sum_{i=1}^{N(t)} X_i = \sum_{i=0}^{N(t)} X_i,
\end{displaymath}

where:
\begin{itemize}
    \item $t$ is a total number of rounds in a single simulation.
    \item Cascades sizes $X_i : i \geq 0$ are independent random variables from
          a distribution generated by \emph{model $\alpha^k$}.
    \item $N(t) \sim \operatorname{Pois}(t\lambda)$ is a \emph{Poisson process} with rate $\lambda$.
\end{itemize}

Such definition of $Y(t)$ is called \emph{compound Poisson process}. Because we
always start with the initial cascade, \emph{multi-source model $\alpha^k$} is a simple extension of
this process.

\begin{figure}[ht!]
    \centering
    \includegraphics[width=0.5\textwidth]{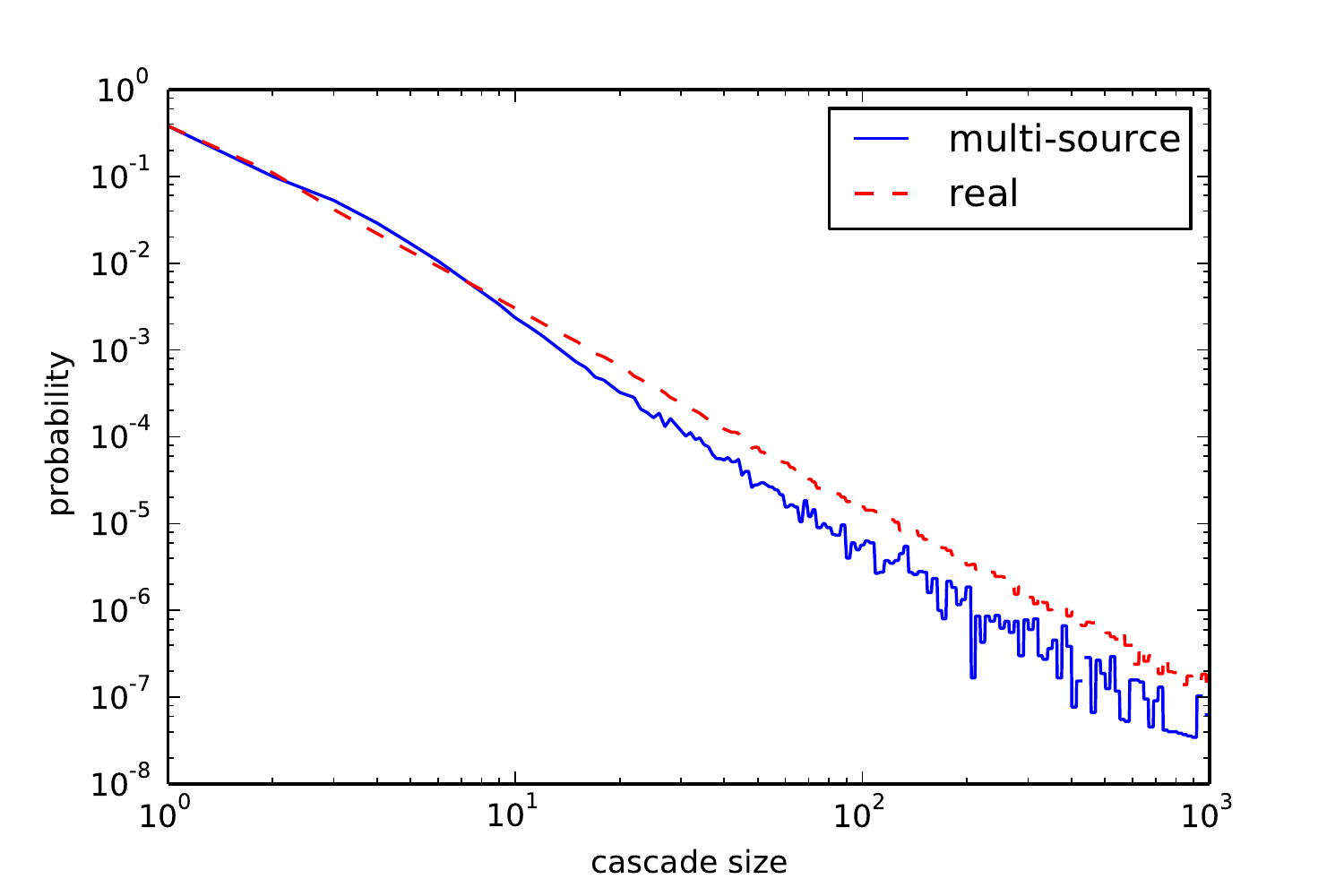}
    \caption{The cascade sizes distribution for multi source $\alpha^k$.
    The data was bucketized into buckets of logarithmic size to reduce variance
    (as suggested in~\cite{log-bucket})}
    \label{fig:model_multisource_dist}
\end{figure}

\subsubsection{Experimental Simulations}

Based on the Algorithm~\ref{multisource_alpha_k_algorithm}, we simulated
\emph{multi-source model $\alpha^k$} on the Twitter retweets graph.
We conducted an effective \emph{grid search} algorithm over parameters $\alpha$
and $\lambda$ with step size $0.0001$.
For each pair of parameters we run $10$ million simulations to obtain a cascade size
distribution. Finally, for them, we computed K-S test.
The minimal value of K-S test $0.0116$ was obtained for $\alpha=0.1215$ and
$\lambda=0.1850$. The K-S test value is roughly the half of the value for model $\alpha^k$
The results for $\lambda=0.0$ are the same as the results
of model $\alpha^k$, since $\lambda=0.0$ implies that we generate only a single source of an information.

On Figure~\ref{fig:model_multisource_dist} one can see the comparison of the
real distribution of cascades sizes and the distribution generated by multi-source model $\alpha^k$ for the best pair of parameters $\alpha$ and $\lambda$.


\section{Discussion and Future work} \label{conclusion}

The observation that rumor popularity follows the power-law enables researchers
to precisely model the information diffusion in the social network. Here, we
showed that simple, one parameter models are insufficient. These models produce
cascades either extremely large or small. That phenomenon is known as the
\emph{phase transition}~\cite{cikm2014}. Gradually decreasing infectiousness of
the information  over time prevents this problem and results in more accurate
predictions.

Our main contribution is that the rumor can spread through a different, unknown media.
We proposed to model it by informing random nodes in the network. This
improvement significantly boost the estimated distribution of cascade size.
Finally, it demonstrates that the underlaying network of the social interactions is much
more complex than just the graph of retweets and the study on the new ways of
estimating it needs more attention. 

Based on the observations made by \citet{leskovec-blog}
regarding the cascade collisions, we proposed the method based on the \emph{compound
Poisson process}. This enables us to produce accurate multi-sourced cascades
with a very little cost. Moreover, this technique significantly lowered the simulation time.

\begin{figure}[ht!]
    \centering
    \includegraphics[width=0.5\textwidth]{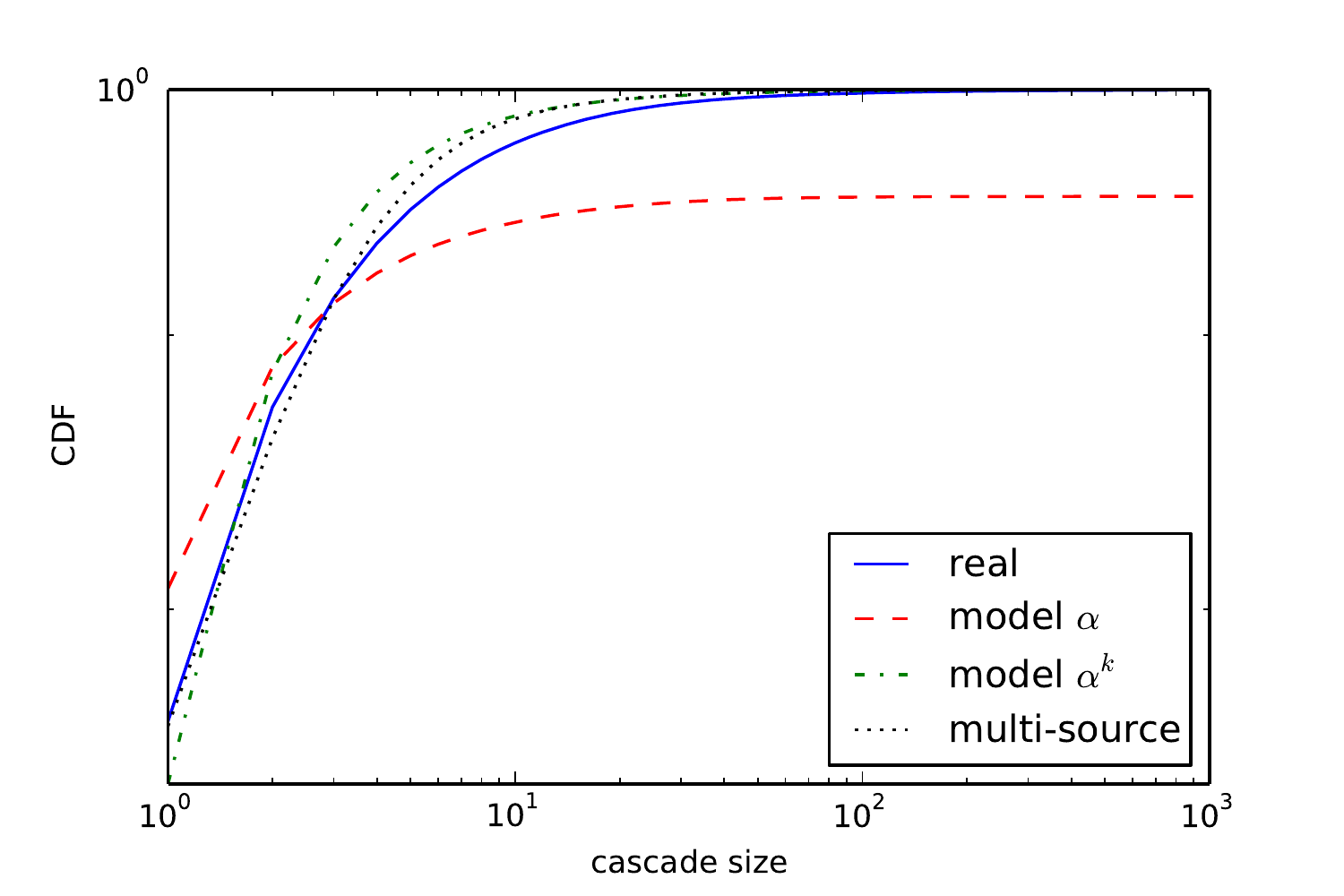}
    \caption{The comparison of CDFs for the discussed models.}

\label{fig:models_cdf}
\end{figure}

We present the final results and comparison of these models in
Table~\ref{tab:final_results}. On Figure~\ref{fig:models_cdf} 
we have shown the comparison between the CDFs
computed for different models. 
Recall, that the K-S test responds to the maximal difference of the CDFs. The
best K-S test value is obtained for the multi-source $\alpha^k$ model.  In order to speed up
simulation of model $\alpha$, we have truncated simulations of $4\%$ of largest
cascades.  Cutting off these cascades does not change the final K-S test
results, because in the real cascade size distribution the probability that the
size of a cascade is larger than $1\,000$ is less than $0.001$.

\begin{table}[ht!]
    \centering
    \caption{The K-S test comparison of the discussed models with the real cascade size
    distribution (log-log scale).}
    \begin{tabular}{|l|l|}
        \hline Model                                 & K-S test \\
        \hline \emph{Model $\alpha$}                 & 0.0447 \\
        \hline \emph{Model $\alpha^k$}               & 0.0207 \\
        \hline \emph{Model multi-source $\alpha^k$}  & 0.0116 \\
        \hline
    \end{tabular}
\label{tab:final_results}
\end{table}

To obtain even more accurate results, one would need to incorporate more complex
effects, for example:

\begin{itemize}
    \item Geographically close nodes might be informed
        through an unknown social network. Close nodes should be informed
        with higher probability than distant~\cite{geo}.
    \item The probability of randomly informing a node may decrease in time because
        the information may become obsolete~\cite{cikm2014}.
    \item The evolution of the social network structure within time~\cite{evolution}.
\end{itemize}

As mentioned in Section~\ref{dataset}, the probability of spreading the
information (i.e., parameter $\alpha$ in the model $\alpha$) decrease with the
size of the graph of retweets. It would be very interesting to investigate the
dependence between the model parameters and the density of the graph of
retweets. Moreover, it would be interesting to study the growth rate of the
graph of retweets.

We have published the gathered data~\cite{approx-data-2016} and the efficient
source code for simulating the basic models~\cite{approx2016}.

\section{Acknowledgments}

This work was supported by ERC StG project PAAl 259515 and FET project MULTIPLEX
317532. We would also want to thank Stefano Leonardi for sharing the data and Rafa{\l} Lata{\l}a
for meaningful discussions.

\bibliographystyle{plainnat}
\bibliography{bib}


\end{document}